\documentclass[conference]{IEEEtran}
\IEEEoverridecommandlockouts

\usepackage{etoolbox}
\usepackage{xstring}
\usepackage{booktabs}
\usepackage{hyperref}

\DeclareListParser{\doslashlist}{/}
\newcounter{ndnNameComponentCounter}%
\newcommand{\Name}[1]{{%
  \setcounter{ndnNameComponentCounter}{0}%
  \renewcommand{\do}[1]{{%
    \ifnumgreater{\value{ndnNameComponentCounter}}{0}{\allowbreak/}{}%
    \ifnumodd{\value{ndnNameComponentCounter}}{}{}%
    \detokenize{##1}}%
    \stepcounter{ndnNameComponentCounter}}%
``{\fontfamily{cmtt}\selectfont\IfBeginWith{#1}{/}{/}{}\doslashlist{#1}}''%
}}

\usepackage{todonotes}
\usepackage{tikz}
\usepackage{amsmath}

\usepackage{filecontents}

\usepackage{tikz}
\usepackage{textcomp}
\usepackage{hyperref}
\usepackage{lipsum}

\newcommand\copyrighttext{%
  \footnotesize \textcopyright 2024 IEEE. Personal use of this material is permitted.
  Permission from IEEE must be obtained for all other uses, in any current or future
  media, including reprinting/republishing this material for advertising or promotional
  purposes, creating new collective works, for resale or redistribution to servers or
  lists, or reuse of any copyrighted component of this work in other works.
  DOI: \href{https://doi.org/10.1109/SCW63240.2024.00108}{10.1109/SCW63240.2024.00108}}
\newcommand\copyrightnotice{%
\begin{tikzpicture}[remember picture,overlay]
\node[anchor=south,yshift=10pt] at (current page.south) {\fbox{\parbox{\dimexpr\textwidth-\fboxsep-\fboxrule\relax}{\copyrighttext}}};
\end{tikzpicture}%
}

\begin{document}

\title{LIDC: A Location Independent Multi-Cluster Computing Framework for Data Intensive Science}

\author{\IEEEauthorblockN{Sankalpa Timilsina}
\IEEEauthorblockA{\textit{Computer Science Department} \\
\textit{Tennessee Tech}\\
Cookeville, TN \\
stimilsin43@tntech.edu}
\and
\IEEEauthorblockN{Susmit Shannigrahi}
\IEEEauthorblockA{\textit{Computer Science Department} \\
\textit{Tennessee Tech}\\
Cookeville, TN \\
sshannigrahi@tntech.edu}
}
\maketitle
\copyrightnotice

\begin{abstract}

Scientific communities are increasingly using geographically distributed computing platforms. The current methods of compute placement predominantly use logically centralized controllers such as Kubernetes (K8s) to match tasks to available resources. However, this centralized approach is unsuitable in multi-organizational collaborations. Furthermore, workflows often need to use manual configurations tailored for a single platform and cannot adapt to dynamic changes across infrastructure.

Our work introduces a decentralized control plane for placing computations on geographically dispersed compute clusters using semantic names. We assign semantic names to computations to match requests with named Kubernetes (K8s) service endpoints. We show that this approach provides multiple benefits. First, it allows placement of  computational jobs to be independent of location, enabling any cluster with sufficient resources to execute the computation. Second, it facilitates dynamic compute placement without requiring prior knowledge of cluster locations or predefined configurations.

\end{abstract}

\section{Introduction} \label{introductionChapter}

The field of scientific research is witnessing a growing demand for compute-intensive tasks. Researchers typically carry out these tasks on local computer clusters, shared community resources, or cloud platforms, each of which presents its own challenges. The cloud incurs high costs, 
and while institutional and shared community clusters are generally well-suited for small-scale scientific workflows \cite{foster2001anatomy}, they remain difficult to locate and utilize, vary in their capabilities, and often require user accounts and configurations specific to the platform.

These institutional and shared compute clusters also operate in isolation. Users must first identify which compute cluster can handle their workflow \cite{sahni2016workflow}, obtain the necessary permissions, create individual user accounts, and manually configure workflows to specify resource requirements such as the number of CPUs and memory. As a result, users are unable to place compute jobs universally and must tailor their workflows to specific platforms.

The scientific community has attempted to mitigate these issues through orchestration systems like Kubernetes (K8s) that can manage resources and place computations in a platform agnostic way. Tools such as Virtual Kubelet and Cilium Mesh extend Kubernetes' capability to handle workloads across multiple clusters. However, these tools depend on complex configurations, specific Container Network Interfaces (CNIs), and manual setup processes \cite{autotuning_in_hpc_applications}. More critically, they still rely on a logically centralized control plane, managed by a central entity. This model struggles to handle dynamic cluster environments or integrate seamlessly across heterogeneous infrastructures. Additionally, users must continue tailoring their workflows to specific clusters and adapt when infrastructure changes \cite{sahni2016workflow}.

Our work directly addresses these challenges in the current compute placement model. Rather than using a central controller, we propose a name-based framework, Location Independent Data and Compute (LIDC), that creates and utilizes a decentralized control plane at the network layer. Our framework creates a loosely coupled overlay of compute clusters using named cluster endpoints. User applications use semantic names for computation jobs  before sending those requests into the network. Such semantically meaningful names also capture job details such as job types and resource requirements. Named Data Networking (NDN) primitives at the network layer forward these compute requests to named cluster endpoints, and once they reach a cluster, LIDC matches compute jobs with a named K8s service endpoint that actually performs the computation. By integrating name-based routing both in the network and within Kubernetes, we enable a seamless end-to-end job placement with LIDC.
LIDC demonstrates that name-based semantics effectively place computation across multiple clusters. Data for these computations are also annotated with names, allowing compute platforms to retrieve raw datasets from a data lake and publish intermediate datasets back to the lake \cite{wang2023gnsga,presley2024hydra}.

This approach enables dynamic and transparent computation placement without requiring users to know cluster locations or configurations. By annotating the requirements and offloading the matchmaking process to the network and the service end-points, LIDC supports seamless job placement, addition and removal of clusters in the compute overlay, and helps create a more flexible and resilient computation overlay that adapts in real-time to changes in load, network conditions, or cluster availability.

\section{Location-Independent Compute Design} 
The LIDC framework allows researchers to make computation requests across any available clusters. LIDC eliminates the need to locate a particular cluster, be familiar with the cluster-specific details, and create cluster-specific configurations by providing a location-transparent interface. The location transparency is achieved using semantically meaningful names for both computational jobs and service endpoints. 

\begin{figure}[!t]
    \includegraphics[width=0.8\linewidth]{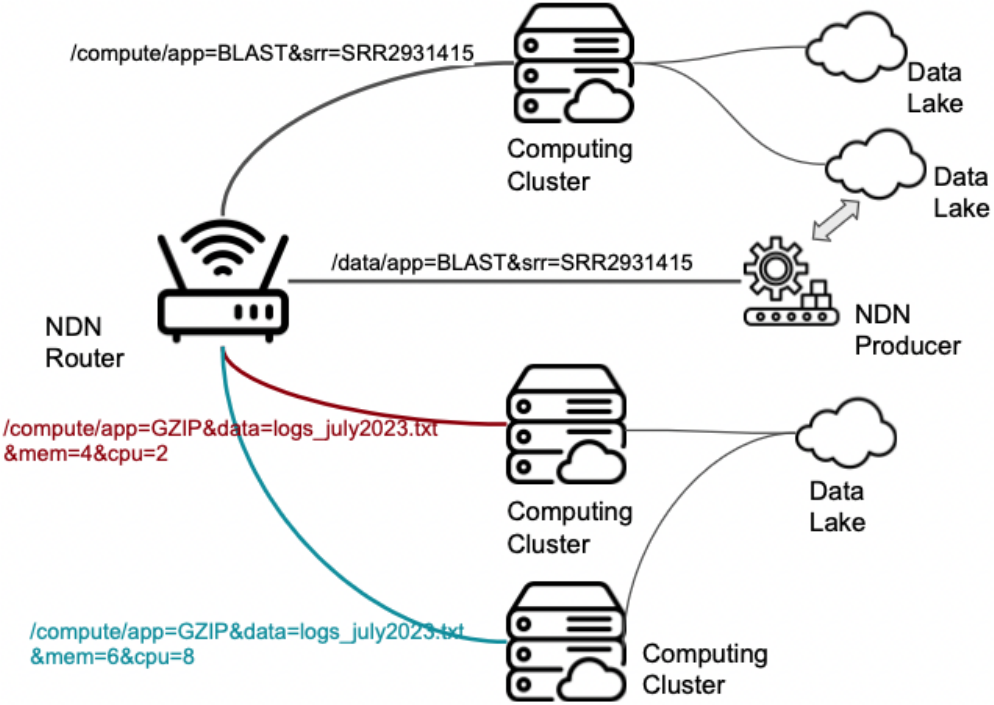}
        \caption{Design of Compute Framework with NDN}
    \label{fig:compute_design}

\end{figure}

Figure \ref{fig:compute_design} illustrates this approach. The semantically named request describes the computation task the user intends to perform and the requested resources. The framework then chooses an appropriate cluster and executes computation tasks based on various factors such as the applications being served by different clusters, resource requirements (such as memory and CPU), past performances, caching, and load balancing capabilities. The framework uses Named Data Networking (NDN) primitives to forward requests to the appropriate clusters. Since the execution of computations requires processing initial or intermediate datasets, the framework also  integrates data lakes built-upon content names for publishing and retrieving datasets.

While this work utilizes NDN primitives for naming computations and employs K8s named service endpoints for actual compute placement, the framework does not need to be tied to these specific technologies. Any orchestration platform, whether currently available or built in the future, can be used to create this framework. Similarly, HTTP(s)-based naming of computational jobs can also match them to appropriate endpoints.

\section{LIDC: Components} \label{lidc:components}
In this section, we describe the components of the LIDC framework. 

\subsection{Kubernetes for Compute Orchestration} \label{lidc:kubernetes_and_its_role}

As we mentioned earlier, we utilize K8s for the compute placement within a cluster. We used K8s because it is the most widely used orchestration system, but other such systems can also be used in its place. 

The integration with Kubernetes serves several distinct purposes:

\begin{itemize}
    \item Naming Computations and Service Endpoints:  Kubernetes' built-in DNS allows for the use of static DNS names that follow established naming conventions for accessing custom applications/service endpoints. This feature also enables us to customize the external endpoints of Kubernetes, essentially the applications Kubernetes exposes outside its cluster. As a result, we can use semantic names to match computational jobs to the applications based on names. 

    \item Scalability:  Kubernetes provides the ability to scale horizontally and vertically, allowing adjustments to the number of applications and platform resources such as memory and CPU. This approach reduces the need for frequent manual interventions to tune computing platforms \cite{autotuning_in_hpc_applications}. Once the resources are appropriately allocated, Kubernetes handles performance degradation or failures, meaning that the network can only serve as a simple matchmaker between services and jobs rather than a complete orchestration system.
\end{itemize}


\subsection{Naming Computation and Services}
\begin{figure}
    \includegraphics[width=\linewidth]{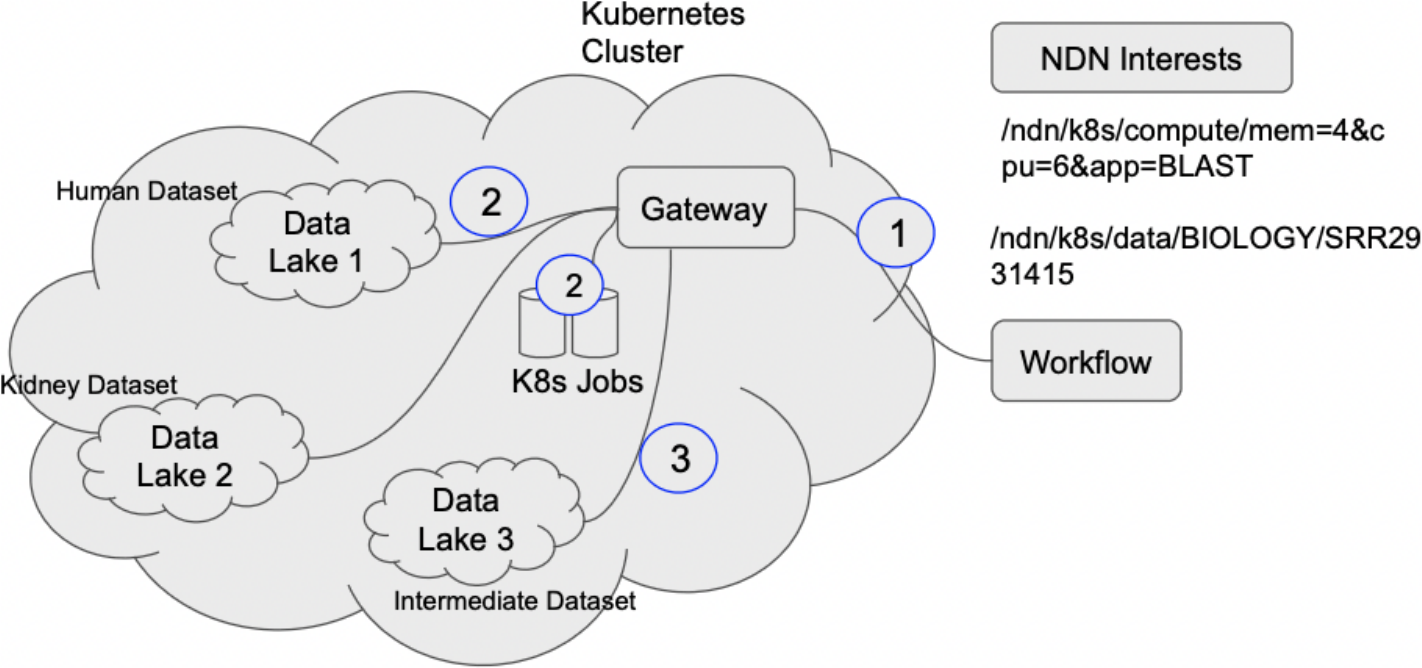}
        \caption{Transparent Data and Compute Placement Based on Names}
    \label{fig:data-compute-bridge}

\end{figure}

The applications running inside Kubernetes, such as a file server, can be accessed using a static Kubernetes DNS name. Such naming is made possible through a Kubernetes service, which is a logical abstraction for a group of deployed pods in a cluster, all of which perform the same function. When a Kubernetes service is created, it is assigned a DNS name within its namespace, which resolves to the IP address of a pod (the smallest execution unit inside Kubernetes) where the application is running. 

When named computing jobs are sent to clusters, they can be directly linked to an application that serves the job. Since we use NDN Interests to carry the jobs to the cluster, we can directly map it to a static DNS name. This DNS name points to a local NDN forwarder, sending it to an appropriate named service within K8s.

The ability to name services and jobs means that the originating workflows do not need to know any internal details of Kubernetes clusters. Simply expressing a request to the externally exposed point will take the request to the underlying service. The utilization of names between workflows and computation clusters is very powerful --- it removes the necessity to manually locate and place computation at a particular compute cluster. Indeed, if multiple clusters expose the same service over an NDN network, the network can bring the compute request to the nearest (or the best) compute cluster.

\subsection{Bridging Data and Computation}

\begin{figure}[!t]
    \includegraphics[width=0.8\linewidth]{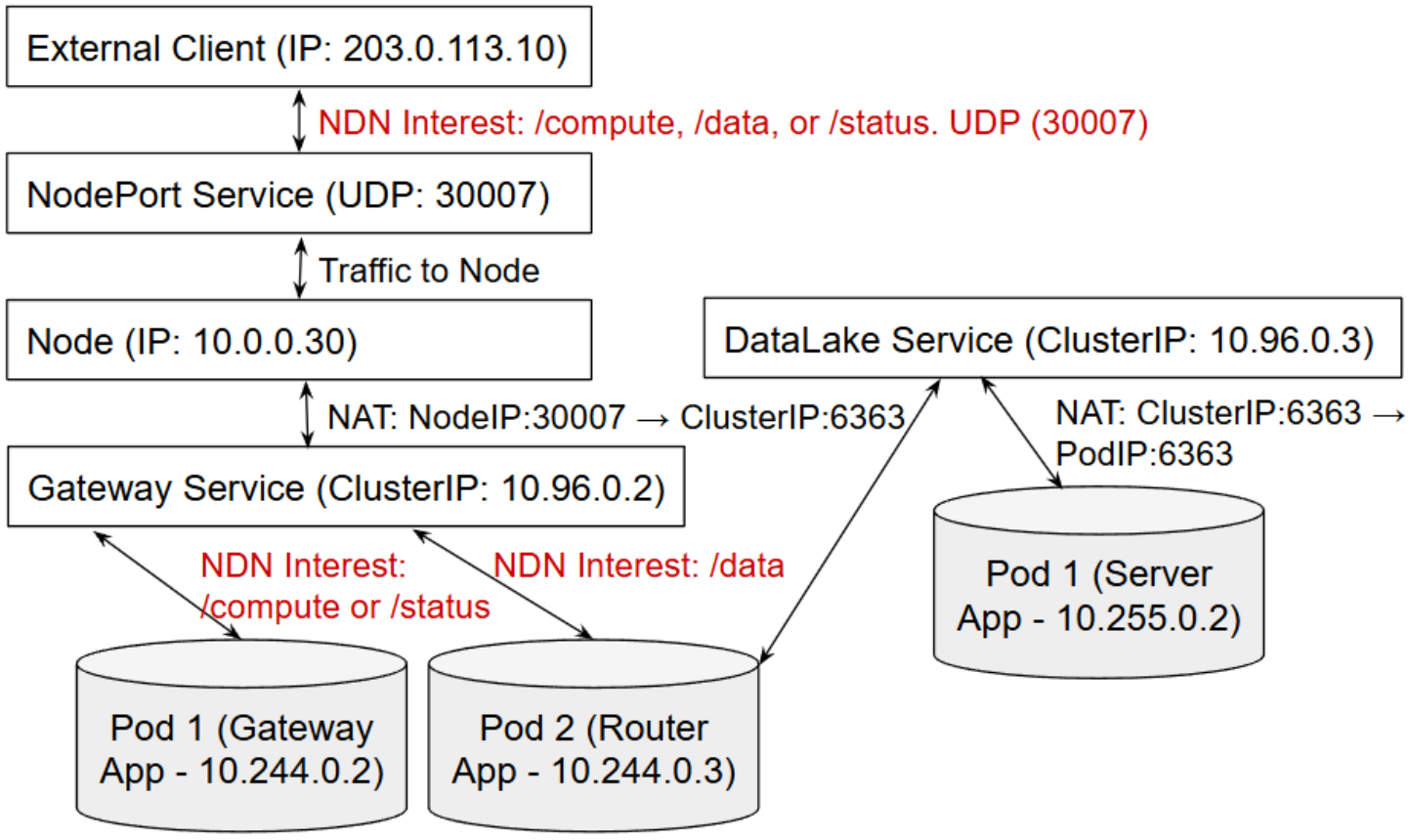}
        \caption{Mapping LIDC to K8s Components}
    \label{fig:client_k8s_flow}
\end{figure}

Figure \ref{fig:data-compute-bridge} shows the high-level constructs of how data and computation are placed together within a cluster. LIDC supports multiple clusters as well.

A workflow uses an NDN Interest name to describe a compute and/or data request, detailing both elements that is needed for specific computations.  These names conveniently tie together the computation and data. An example of such name might be ``/ndn/k8s/compute/\texttt{<}compute-name\texttt{>}\&\texttt{<}compute-parameters\texttt{>}\&\texttt{<}dataset-names\texttt{>}''. A practical representation of this naming pattern can be ``/ndn/k8s/compute/mem=4\&cpu=6\&app=BLAST'', as Figure \ref{fig:data-compute-bridge} shows.

Figure \ref{fig:client_k8s_flow} shows how these requests map to existing K8s components. 
External clients can connect to a Kubernetes service using a NodePort. When using NodePort, the Kubernetes control plane automatically assigns a port from a specified range (e.g., 30000-32767) and makes the service accessible on each node's IP address at that port. In our setup, we expose the Gateway's NFD application as one of the Kubernetes services. This means that an NDN client outside the cluster can establish a direct socket-based connection to the exposed port and IP address of the NFD application.

\begin{figure}
    \includegraphics[width=\linewidth]{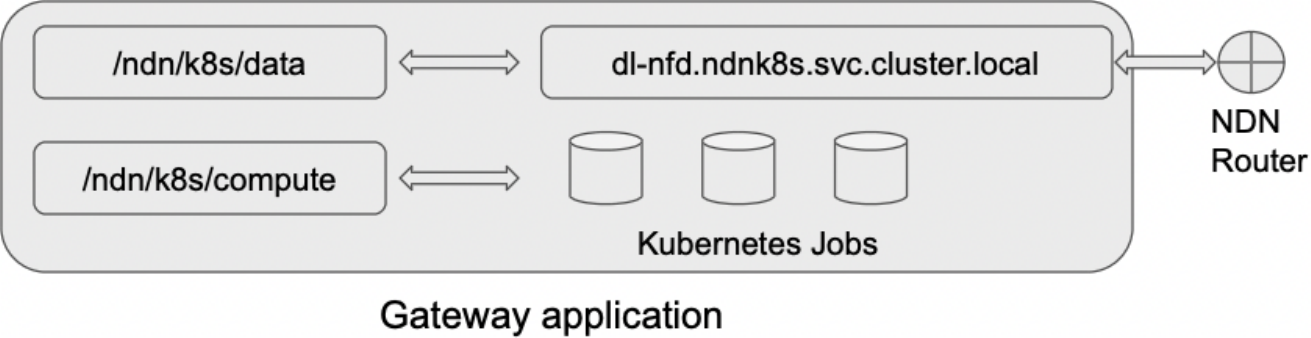}
        \caption{Mapping NDN names to Kubernetes services}
    \label{fig:computation-k8s-mapping}
\end{figure}

Figure \ref{fig:computation-k8s-mapping} shows a custom gateway application running on a Kubernetes cluster. After connecting to the cluster, external clients can send Interests to the exposed Gateway application. The Gateway acts as a decision-maker, determining how to process the incoming Interest. If the Interest relates to computational tasks, the Gateway parses the Interest to understand details such as the specific application to be activated, the target dataset, and other application parameters like memory capacity and CPU needs. Once these details are clear, the Gateway initiates a Kubernetes job to run the desired computation task.


Internally, one of the applications deployed on LIDC is an NDN router \cite{nfddev}. This application can be accessed via a Kubernetes DNS, for example, ``dl-nfd.ndnk8s.svc.cluster.local". This router serves as a gateway to various internal applications, including a data lake (which serves data under `/ndn/k8s/data') and a file server that provides Genomics files \cite{iordache2024named}. 

Once 
the
computations are done, the intermediate results can be stored back in the same data lakes. For future access or reference, one can easily retrieve these results by sending a standard data request, like ``/ndn/k8s/data/\texttt{<}data-identifier\texttt{>}'', back to the same cluster.

LIDC is an open-source implementation and is publicly available at \cite{ndn_with_kubernetes}. By default, the LIDC is setup with a single Kubernetes node. This node is the gateway to the cluster and is the only node that can be accessed. 

\section{Deploying a Genomics Workflow using LIDC}
\begin{figure}

    \includegraphics[width=0.8\linewidth]{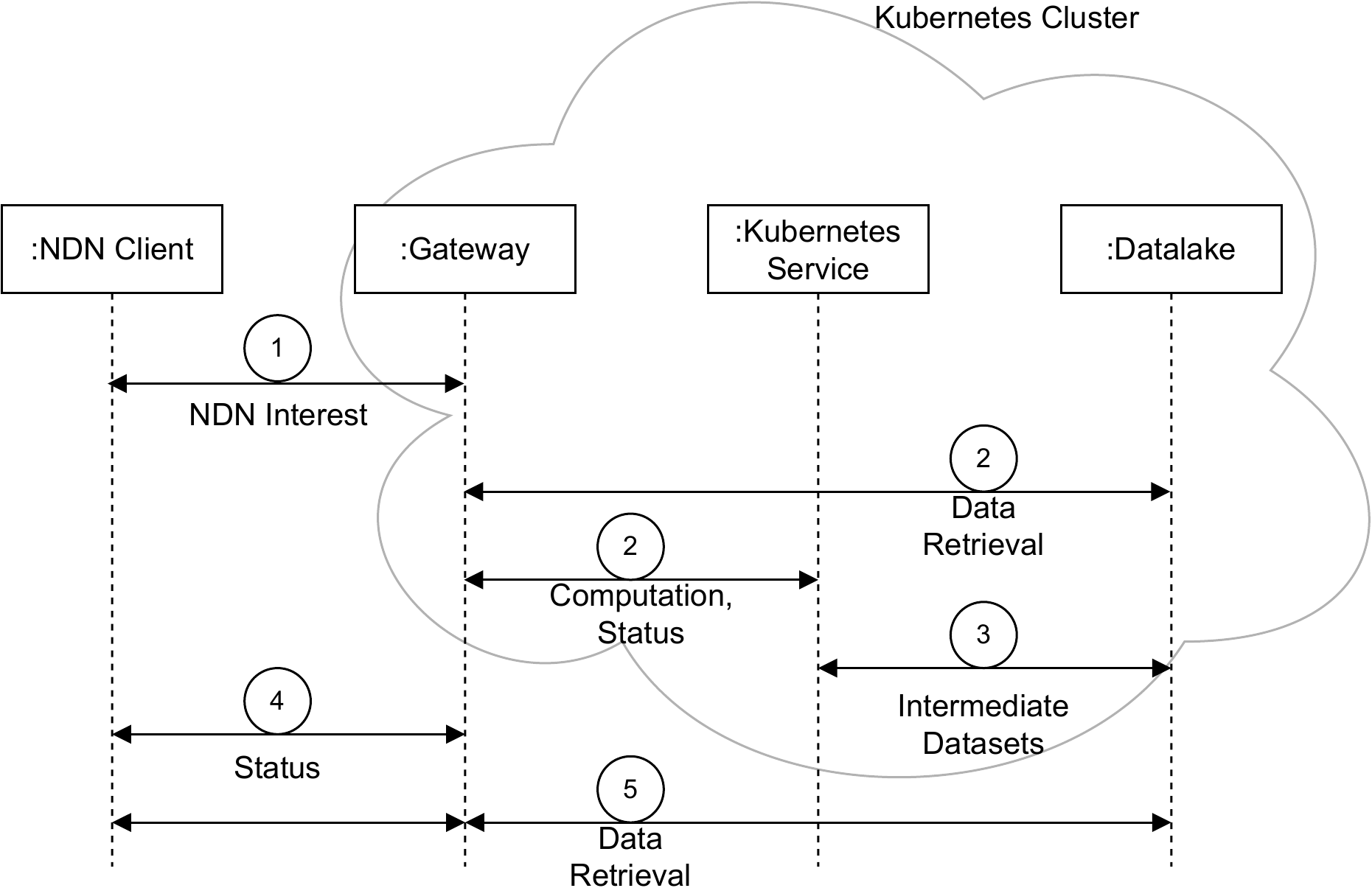}
        \caption{LIDC: Workflow Details}
    \label{fig:lidc_workflow_details}
\end{figure}

This section presents a real-world example of how LIDC can facilitate the deployment of an actual Genomics workflow.  Figure \ref{fig:lidc_workflow_details} shows the protocol details. 

This particular deployment is equipped with NCBI's Magic-BLAST application \cite{ncbi_magicblast}, a tool that aligns genetic sequences, helping scientists match and compare DNA samples. While we use BLAST, LIDC can incorporate any application. 

Upon setting up the cluster, LIDC configures the following components: (a) a gateway, in which a single NFD pod acts as the gateway to the services running on this cluster, and (b) a Kubernetes PVC (Persistent Volume Claim) and mounts it to an NFS server, which functions like a remote data lake. This is where predefined genomics datasets are downloaded and accessed through the ``/ndn/k8s/data" namespace.

The gateway NFD has prefix registrations for ``/ndn/k8s/data," pointing to the data lake's NFD, and ``/ndn/k8s/compute," which the gateway node itself handles through Kubernetes jobs.

The incoming Interest can have one of the following above two prefixes: (a) ``/ndn/k8s/data": This indicates a request for data from the data lake. Upon receiving the Interest, the gateway NFD will route the request to the data lake's NFD. The data lake's NFD is complemented by a fileserver application, which serves the data from the PVC. (b) ``/ndn/k8s/compute": This prefix indicates a request for computation. The Interest is first parsed on the gateway node to understand the computing requirements. The gateway node then runs a Kubernetes job with the specified resources. The gateway can directly spawn a new computation task at runtime. The computation job can use data directly from the data lake if needed, and any intermediate or final datasets are stored back in the data lake.

The client can inquire about the status of a job by asking the gateway, which then checks with the Kubernetes service and responds. After learning that the task is done, the client can retrieve the data from the data lake through the gateway.

\subsection{Interests for named computations}
The framework provides a sample client application \cite{ndn_with_kubernetes}, to make computation requests. An example of this is a client asking to BLAST \cite{ncbi_magicblast} a known SRR\textunderscore ID against a human genome reference dataset. The application parameters, in this case like SRR\textunderscore ID, memory and cpu requirements are encoded in the NDN Interest itself by the client application.

Besides the ``/ndn/k8s/data'' and ``/ndn/k8s/compute'' NDN prefixes mentioned in  the previous section, the framework also offers a ``/ndn/k8s/status'' prefix for clients to perform status checks of their computation tasks. Clients need a \emph{job\textunderscore id} from their initial ``/ndn/k8s/compute'' request to use this. When checking the status, the LIDC responds with one of the following states:

\begin{itemize}
    \item Completed: The application has completed running. The response contains the information as to how to retrieve the results from the data lake.
    \item Failed: The application has errored. The response contains error message.
    \item Running: The application is running.
    \item Pending: The application is starting.
\end{itemize}

\subsection{Application Specific Validations}
Finally, LIDC allows for application-specific validations. These validations are built into the system in a modular manner and can be managed separately for each application. For instance, with Magic-BLAST \cite{ncbi_magicblast}, a specific check might be confirming correct SRR\textunderscore IDs in the provided Interest. Another application, like a file compression tool, might not need SRR\textunderscore IDs and could have its own checks. These checks can be set up individually for each application.

\section{Testbed Deployment}
For this work, we used Google Cloud Platform (GCP) virtual machines to set up the Kubernetes cluster. This section describes a lightweight Kubernetes distribution used to setup the cluster and the specifics of publication and data retrieval.

\begin{table*}
\centering
    \caption{Computation Performance}
    \label{table:computation_perf}
    \begin{tabular}{@{}p{2.5cm}p{2cm}p{1.5cm}p{1.5cm}p{1cm}p{2cm}p{2cm}@{}}
        \toprule
        SRR\textunderscore ID & Ref. Database & Genome Type & Memory (GB) & CPU & Run Time & Output Size\\
        \midrule
        SRR2931415 & HUMAN & RICE & 4 & 2 & 8h9m50s & 941MB\\
        \midrule
        SRR2931415 & HUMAN & RICE & 4 & 4 & 8h7m10s & 941MB\\
        \midrule
        SRR5139395 & HUMAN & KIDNEY & 4 & 2 & 24h16m12s & 2.71GB\\
        \midrule
        SRR5139395 & HUMAN & KIDNEY & 6 & 2 & 24h2m47s & 2.71GB\\
        \bottomrule
    \end{tabular}
\end{table*}
\subsection{MicroK8s on GCP}
MicroK8s \cite{microk8s} is a streamlined Kubernetes distribution optimized for developers, edge computing, IoT, and small-scale deployments. Its advantage lies in its simplicity, stripping away the intricacies often tied to large-scale, multi-node clusters. For this work, we leveraged Google Cloud Platform (GCP) to set up MicroK8s on a single virtual machine (VM). To bolster the communication and service discovery mechanisms within our cluster, we enabled the DNS add-on, which employs CoreDNS to provide address resolution services specific to Kubernetes. Given that this DNS service often underpins the operation of other add-ons, its activation is crucial to enable service discovery using DNS names we described in earlier sections.

For our tests, we installed MicroK8s with a Network File System (NFS) server. Next, we downloaded and setup the human reference database and set up rice and kidney sample Sequence Read Archive files from NCBI \cite{ncbi_sra}. This setup was automated using a script and is available at \cite{ndn_with_kubernetes}. The NFS server is analogous to a remote file server. We mount this to the running MicroK8s Kubernetes cluster with declarative configurations.

\subsection{Creating and Loading PVCs}
To evaluate the crucial step of creating and loading the PVCs of the data lake with content to be published, LIDC provides data loading tool \cite{ndn_with_kubernetes} that downloads and sets up human reference database and sample Sequence Read Archive (SRA) genome files. We loaded two genomic datasets: a 36-samples of sequenced human kidney tumor RNA \cite{SRA_kidney}, and a 99 samples of sequenced rice RNA \cite{SRA_rice}. Each dataset was loaded onto two independent PVCs. This is an one time operation and does not contribute to the overall delay in subsequent data retrieval operations.





\section{Results}
As a proof-of-concept, we compared sample SRA experiments with a human reference database to identify similar genome sequences. This produces a compressed file from the comparison. We recorded the input setup, computation time, and output file size. This is shown in the Table \ref{table:computation_perf}.


We BLASTed \cite{ncbi_magicblast} sample rice and kidney datasets with a human reference database. The computation was carried out with different CPU and memory configurations, and we noted down the total run time. The takeaway from this result could be how different configurations are impacting the run time. In our specific case, a variance of CPU and memory sizes is not showing any significant changes in the run time. If we deploy intelligence in the network, then the network can learn from this data and be able to pick the optimal configuration for future tasks.

\section{Discussion and Future Work} \label{sec:discussion_and_future_work}
The work introduces a framework for computation that is not tied to a specific location or underlying infrastructure. This capability means that the network can dynamically identify and utilize resources based on current conditions, eliminating the need for manual interventions in cluster management. 

The LIDC system enhances flexibility without significantly increasing potential attack points. NDN inherently secures data and provides built-in data authentication and integrity. Kubernetes includes security features such as role-based access control (RBAC) and network policies, which can be integrated with attributes based encryption \cite{reddick2022wip}. By decentralizing control, LIDC reduces the risks associated with a single point of failure and compromising a central controller. Additionally, our approach uses semantic names instead of exposing cluster locations or configurations, which may reduce potential attack surfaces. However, further research is needed to create a security framework for multi-cluster compute placement.

In the future, we plan to enhance the control plane by introducing intelligence. For example, we aim to enable the network to identify the most suitable cluster for executing requests and optimize the system by leveraging machine learning algorithms to predict completion times. Once the network knows cluster capabilities, it can select the best cluster based on computing and timing requirements, data size, past performances, and other factors.

Additionally, implementing result caching in the framework would be beneficial, primarily when multiple clients issue identical requests. This can be achieved by uniquely identifying names and using various storage solutions such as databases, file storage, or object storage to store the mapping information.


\section{Conclusion}
While orchestration systems such as Kubernetes have made significant strides in multi-cluster management and orchestration, our proposed framework offers a fundamentally different approach by leveraging Named Data Networking (NDN) to achieve true location independence, seamless data and compute integration, and automated multi-cluster orchestration. This novel approach addresses several limitations and complexities inherent in Kubernetes-based solutions, such as manual configuration and adaptation to dynamic changes, particularly in data-intensive scientific workflows. By embedding compute placement capability directly into the network layer, the framework provides a more flexible, scalable, and user-friendly platform that is well-suited to the evolving needs of the scientific community.




\section*{Acknowledgments}
This material is based upon work supported by the National Science Foundation under Grant No. \#2126148.

\bibliographystyle{plain}
\bibliography{references}


\end{document}